\documentclass[11pt]{article}

\usepackage{amssymb,amsmath,amsfonts,amsthm}
\usepackage{graphicx}

\usepackage{hyperref}
\input xy
\xyoption{all}

\def\be{\begin{equation}}
\def\ee{\end{equation}}
\def\bea{\begin{eqnarray}}
\def\eea{\end{eqnarray}}

\newcommand{\sect}[1]{\setcounter{equation}{0}\section{#1}}
\newcommand{\subsect}[1]{\subsection{#1}}

\renewcommand{\theequation}{\arabic{section}.\arabic{equation}}

\theoremstyle{plain}
\newtheorem{theorem}{Theorem}

\newtheorem{proposition}[theorem]{Proposition}

\theoremstyle{definition}

\numberwithin{theorem}{section}
\numberwithin{equation}{section}

\def\dd{{\rm d}}
 \def\lh{{\cal H}_\omega}
  
  \def\qlhzd{   C^\infty({\cal H}^*_{z,\omega}) }

  \def\shc{\,{\rm shc}}
  \def\sinc{\,{\rm sinc}}
  \def\cosh{\,{\rm ch}}
  \def\sinh{\,{\rm sh}}
  \def\tanh{\,{\rm th}}

\def\1{\'{\i}}

\def\eee{{\rm e}}

 \def\xx{u}
 \def\yy{v}

\begin{document}

\thispagestyle{empty}

\
\medskip
\medskip

 \vskip2cm

 \begin{center}

\medskip
\medskip

\noindent {\LARGE{\bf {Poisson--Hopf algebra deformations\\[6pt] of  Lie--Hamilton  systems}}}  

\medskip
\medskip

\medskip
\medskip
\medskip

{\sc  \'Angel Ballesteros$^{1}$, Rutwig Campoamor-Stursberg$^{2,3}$, Eduardo Fern\'andez-Saiz$^{3}$,\\[4pt]   Francisco J.~Herranz$^{1}$ and  Javier de Lucas$^{4}$}

 \end{center}

\medskip

\noindent
$^1$ Departamento de F\1sica, Universidad de Burgos,   E-09001 Burgos, Spain

\noindent
$^2$ Instituto de Matem\'atica Interdisciplinar I.M.I-U.C.M., E-28040 Madrid, Spain

\noindent 
$^3$ Departamento de Geometr\1a y Topolog\1a,  Universidad Complutense de Madrid, Plaza de Ciencias 3, E-28040 Madrid, Spain

\noindent
$^4$ Department of Mathematical Methods in Physics, University of Warsaw, Pasteura 5, 02-093 Warszawa, Poland

\medskip

\noindent
{\small 
 E-mail: {\rm  angelb@ubu.es, rutwig@ucm.es, eduardfe@ucm.es, fjherranz@ubu.es,  	  javier.de.lucas@fuw.edu.pl 
 }}

  \medskip

\begin{abstract}
\noindent 
Hopf algebra deformations are merged with a class of Lie systems of Hamiltonian type, the so-called Lie--Hamilton systems, to devise a novel formalism: the Poisson--Hopf algebra deformations of Lie--Hamilton systems. This approach applies to any Hopf algebra deformation of any Lie--Hamilton system. Remarkably, a Hopf algebra deformation transforms a Lie--Hamilton system, whose dynamic is governed by a finite-dimensional Lie algebra of functions, into a non-Lie--Hamilton system associated with a Poisson--Hopf algebra of functions that allows for the explicit description of its $t$-independent constants of the motion from deformed Casimir functions. We illustrate our approach by considering  the  Poisson--Hopf algebra analogue of the non-standard quantum deformation of  $\mathfrak{sl}(2)$ and its applications to deform well-known Lie--Hamilton systems describing oscillator systems, Milne--Pinney  equations, and several  types of Riccati equations. In particular, we obtain a new position-dependent mass oscillator system with a time-dependent frequency.
\end{abstract}

\bigskip


\medskip


\medskip



\sect{Introduction}

A {\it Lie system} is a nonautonomous
system of first-order ordinary differential equations whose general solution can be written as a function, a so-called {\it superposition rule}, of a family of particular solutions and some constants \cite{LS,VES,DAV}. Superposition rules constitute a structural 
property that emerges naturally from the group-theoretical 
approach to differential equations initiated by Lie, Vessiot, and Guldberg, within the context of 
the development of the geometric program based on transformation groups, as well as from the analytic classification 
of differential equations developed by Painlev\'e and  Gambier,
among others. Indeed, Lie proved that every Lie system can be described by a finite-dimensional Lie algebra of vector fields, a {\it Vessiot--Guldberg Lie algebra} \cite{LS}, and Vessiot used Lie groups to derive superposition rules \cite{VES}. 

In the frame
of physical applications, it was not until the 1980s that the power
of superposition rules and Lie systems was fully recognized \cite{PW}, motivating a systematic
analysis of their applications in classical dynamics and their potential generalization to quantum systems  (see
\cite{PW,CGM00,CGM07,Dissertations} and references therein). 

Although Lie systems, as well as their refinements and generalizations, represent a valuable
auxiliary tool in the integrability study of physical systems, it seems surprising that the methods 
employed have always remained within the limitations of Lie group and distribution theory, without considering 
other 
 frameworks that have turned out to be a very successful
approach to integrability, such as quantum groups and Poisson--Hopf algebras ~\cite{Abe,Chari,Majid,coalgebra1,coalgebra2}. We recall that, beyond superintegrable systems~\cite{coalgebra1,coalgebra2}, Poisson coalgebras have been recently  applied to  integrable bi-Hamiltonian deformations of     Lie--Poisson systems~\cite{Ballesteros1} and to integrable deformations of R\"ossler and Lorenz systems~\cite{Ballesteros2}.

This paper presents a novel generic procedure for the Poisson--Hopf algebra deformations of  {\it Lie--Hamilton (LH) systems}, namely Lie systems endowed with a Vessiot--Guldberg Lie algebra of Hamiltonian vector fields relative to a Poisson structure \cite{CLS13}. LH systems posses also a finite-dimensional Lie algebra of functions, a so-called {\it  LH algebra}, governing their dynamics \cite{CLS13}. Then, our approach is based on the Poisson coalgebra formalism extensively used in the context of superintegrable systems together with the notion of involutive distributions in the sense of Stefan--Sussman (see \cite{Va94,Pa57,WA} for details). The crux will be to consider a  Poisson--Hopf algebra structure 
that replaces the LH algebra of the non-deformed LH system, thus  allowing for an explicit construction of $t$-independent constants of the 
motion, that will be expressed in terms of the deformed Casimir invariants. Moreover, the deformation will generally transform the Vessiot--Guldberg Lie algebra of the LH system into a mere set of vector fields generating an integrable distribution in the sense of Stefan--Sussman. Consequently, the deformed LH systems are not, in general, Lie systems anymore.

Our novel approach is presented in the next section, where the basics of LH systems and Poisson--Hopf algebras are recalled (for details on the general theory of Lie and LH systems, the reader is referred to~\cite{LS,PW,CGM00,CGM07,Dissertations,CLS13, CGL10,CGL11,BCHLS13Ham,BBHLS,BHLS,C132,C135, Ibragimov16,Ibragimov17,HLT}).
To illustrate this construction, we consider in section 3 the Poisson--Hopf algebra analogue of the so-called non-standard quantum deformation of  $\mathfrak{sl}(2)$~\cite{Ohn,beyond, non, Shariati} together with its deformed Casimir invariant. 
 
Afterwards, relevant examples of deformed LH systems that can be extracted from this deformation are given. Firstly, the non-standard deformation of the Milney--Pinney equation is presented in section 4, where this  deformation is shown to give rise to a new oscillator system with a position-dependent mass and a time-dependent frequency, whose (time-independent) constants of the motion are also explicitly deduced.
In sections 5 and 6 several deformed (complex and coupled) Riccati equations are obtained as a straightforward application of the formalism here presented. We would like to stress that, albeit these applications are carried out on the plane, thus allowing a deeper insight in the proposed formalism, the method here presented is by no means constrained dimensionally, and its range of applicability goes far beyond the particular cases here considered. 
Finally, some remarks and open problems are addressed in the concluding section.


\sect{Formalism}
For  the sake of  simplicity we will develop our formalism and its corresponding applications on $\mathbb{R}^2$,  but we stress that this approach can be applied, mutatis mutandis, to construct Poisson--Hopf algebra deformations of LH systems defined on any manifold.


\subsect{Lie--Hamilton systems}

Let us consider the global coordinates  $\{x,y\}$ on the Euclidean plane $\mathbb{R}^2$. Geometrically, every nonautonomous system of first-order differential equations on $\mathbb{R}^2$ of the form
\begin{equation}
 \frac{{\rm d} x}{{\rm d} t  }=f(t,x,y), \qquad \frac{{\rm d} y}{{\rm d} t }=g(t,x,y),
 \label{system}
\end{equation}
where $f,g:\mathbb{R}^3\rightarrow \mathbb{R}$ are arbitrary functions, amounts to a $t$-dependent vector field  ${\bf X}:\mathbb{R}\times \mathbb{R}^2\rightarrow {\rm T}\mathbb{R}^2$  given by 
\begin{equation}\label{Vect}
{\bf X}:\mathbb{R}\times\mathbb{R}^2\ni (t,x,y)\mapsto f(t,x,y)\frac{\partial}{\partial x}+g(t,x,y)\frac{\partial}{\partial y}
\in {\rm T}\mathbb{R}^2 .
\end{equation}
This justifies to represent (\ref{Vect}) and its related system of differential equations (\ref{system}) by ${\bf X}$ (cf. \cite{Dissertations}). Let us assume ${\bf X}$ to be a {\it Lie system} on $\mathbb{R}^2$, namely it admits a superposition rule (see \cite{LS,PW,CGM00,CGM07,Dissertations,CGL10} for details). Since the general solution to a Lie system is not generally known, the use of a superposition rule enables one to unveil its general properties or to simplify the use of numerical methods \cite{PW,Dissertations}. Lie systems are, for instance, several Riccati, Kummer--Schwarz   and Milne--Pinney equations when written as first-order systems of differential equations \cite{Dissertations,BCHLS13Ham, BBHLS,BHLS}.

According to the {Lie--Scheffers Theorem}~\cite{LS,CGM00,CGM07},  a system $\bf X$ is a Lie system if and only if 
\be
{\bf X}_t(x,y):= {\bf X}(t,x,y)=\sum_{i=1}^l b_i(t){\bf X}_i(x,y) ,
\label{aabb}
\ee
  for some $t$-dependent functions $b_1(t),\ldots,b_l(t)$ and vector fields ${\bf X}_1,\ldots,{\bf X}_l$ on $\mathbb{R}^2$ that span  an $l$-dimensional real Lie algebra $V$ of vector fields, i.e. the Vessiot--Guldberg Lie algebra of ${\bf X}$.  

  A Lie system ${\bf X}$ is, furthermore, a LH one~\cite{Dissertations,CLS13,BCHLS13Ham,BBHLS,BHLS, HLT} if it admits a Vessiot--Guldberg Lie algebra $V$ of Hamiltonian vector fields relative to a Poisson structure. This amounts to the existence, around each generic point of $\mathbb{R}^2$, of a symplectic form, $\omega$, such that:
\be
\mathcal{L}_{{\bf X}_i}\omega=0 ,
\label{der}
\ee
for a basis ${\bf X}_1,\ldots,{\bf X}_l$ of $V$ (cf.~Lemma 4.1 in \cite{BBHLS}). To avoid minor technical details and to highlight our main ideas, hereafter it will be  assumed, unless otherwise stated, that the symplectic form and remaining structures are defined globally. More accurately, a  local description around a generic point in $\mathbb{R}^2$ could easily  be  carried out.

Each vector field ${\bf X}_i$  admits a Hamiltonian function $h_i$ given by the rule:
\be
\iota_{{\bf X}_i}\omega={\rm d}h_i,
\label{contract}
\ee
where $\iota_{{\bf X}_i}\omega$ stands for the contraction of the vector field ${\bf X}_i$ with the symplectic form $\omega$. Since $\omega$ is non-degenerate, every function $h$ induces a unique associated Hamiltonian vector field ${\bf X}_h$. This fact gives rise to a Poisson bracket on $C^\infty(\mathbb{R}^2)$ given by
 \begin{equation}\label{LB}
 \{\cdot,\cdot\}_\omega\ :\ C^\infty\left(\mathbb{R}^2\right)\times C^\infty\left(\mathbb{R}^2\right)\ni (f_1,f_2)\mapsto X_{f_2} f_1\in C^\infty\left(\mathbb{R}^2\right),
 \end{equation}
turning $(C^\infty(\mathbb{R}^2),\{\cdot,\cdot\}_\omega)$ into a Poisson algebra \cite{Va94}. The space ${\rm Ham}(\omega)$ of  Hamiltonian vector fields on $\mathbb{R}^2$ relative to $\omega$ is also a Lie algebra relative to the commutator of vector fields. Moreover, we have the following exact sequence of Lie algebra morphisms (see \cite{Va94})
\begin{equation}\label{seq}
0\hookrightarrow \mathbb{R}\hookrightarrow (C^\infty(\mathbb{R}^2),\{\cdot,\cdot\}_\omega)\stackrel{\varphi}{\longrightarrow} ({\rm Ham}(\omega),[\cdot,\cdot])\stackrel{\pi}{\longrightarrow} 0,
\end{equation} 
where $\pi$ is the projection onto $0$ and $\varphi$ maps each $f\in C^\infty(\mathbb{R}^2)$ onto the Hamiltonian vector field ${\bf X}_{-f}$.
In view of the sequence (\ref{seq}), the Hamiltonian functions  $ h_1,\ldots,h_l$ and their successive Lie brackets with respect to (\ref{LB}) span a finite-dimensional Lie algebra of functions contained in $\varphi^{-1}(V)$. This Lie algebra is called a
   {\em  LH  algebra}  $\lh$  of  $X$.  We recall that LH algebras play a relevant role in the derivation of constants of motion and superposition rules for LH systems~\cite{BCHLS13Ham, BHLS,HLT}.


\subsect{Poisson--Hopf algebras}

 The core in what follows is the fact that the space $C^\infty\left(\lh^* \right)$ can be endowed with a {\it Poisson--Hopf algebra} structure. 
  We recall that an associative algebra $A$  with a {\it product} $m$ and a {\it unit} $\eta$  is said to be a {\em Hopf algebra} over $\Bbb R$
\cite{Abe,Chari, Majid} if there exist two homomorphisms called {\em coproduct}  $(\Delta :
A\longrightarrow A\otimes A )$ and {\em counit} $(\epsilon : A\longrightarrow
\Bbb R)$, along with an  antihomomorphism, the {\em antipode} $\gamma :
A\longrightarrow A$,  such that for every $ a\ \! \in A$ one gets: 
$$
\begin{gathered}
({\rm Id}\otimes\Delta)\Delta (a)=(\Delta\otimes {\rm Id})\Delta (a),  
\nonumber \\
({\rm Id}\otimes\epsilon)\Delta (a)=(\epsilon\otimes {\rm Id})\Delta (a)= a, 
\nonumber\\
m((   {\rm Id}  \otimes \gamma)\Delta (a))=m((\gamma \otimes {\rm Id})\Delta (a))=
\epsilon (a) \eta,\nonumber
\end{gathered}
$$
where $m$ is the usual multiplication $m(a\otimes b)=ab$.  Hence the following   diagram is commutative:
 \\
 
   \centerline{ {\xymatrix@C=0.4em{&A\otimes A\ar[rr]^{{\rm Id}\,\otimes\,  \gamma}&&A\otimes A\ar[rd]^m&&\\
   A\ar[rd]^\Delta\ar[rr]^{\epsilon}\ar[ur]^\Delta&&\mathbb{R}\ar[rr]^{\eta}&&A\\
   &A\otimes A\ar[rr]^{\gamma\,\otimes\, {\rm Id}}&&A\otimes A\ar[ru]^m&}	}}

\medskip

\noindent
If $A$ is a commutative Poisson algebra and $\Delta$ is a Poisson algebra morphism, then  $(A,m, \eta,\Delta,\epsilon,\gamma)$ is a {\it Poisson--Hopf algebra}  over $\Bbb R$.
We recall that the   Poisson bracket on $A\otimes A$ reads
$$
\{ a\otimes b, c\otimes d\}_{A\otimes A}=\{ a, c\}\otimes  b d + a c\otimes \{ b, d\} ,\qquad \forall a,b,c,d\in A .
$$

In our particular case, $C^\infty\left(\lh^* \right)$ becomes a Hopf algebra relative to its natural associative algebra with unit provided that
$$
\begin{gathered}
 \Delta (f)(x_1,x_2):=f(x_1+x_2),\qquad m(h\otimes g)(x):=h(x)g(x),\\
 \epsilon (f):=f(0),\qquad \eta(1)(x):=1,\qquad \gamma(f)(x):=f(-x),
 \end{gathered}
 $$
 for every $x,x_1,x_2\in \lh$ and $f,g,h\in C^\infty(\lh^*)$. Therefore, the space $C^\infty\left(\lh^* \right)$ becomes a Poisson--Hopf algebra by endowing it with the Poisson structure defined by the Kirillov--Kostant--Souriau bracket related to a Lie algebra structure on $\lh$.


\subsect{Deformations of Lie--Hamilton systems and generalized distributions}

  The aim of this paper is to provide a systematic procedure to obtain deformations of  LH systems by using  LH algebras and deformed Poisson--Hopf algebras  that lead to appropriate extensions of the theory of LH systems. Explicitly, the construction is based upon the following  {four} steps:

\begin{enumerate}

\item Consider a LH  system ${\bf X}$ (\ref{aabb}) on  $\mathbb{R}^2$ with respect to a symplectic form $\omega$ and admitting a LH algebra $\lh$   spanned by a basis of functions $h_1,\ldots, h_l\in C^\infty(\mathbb{R}^2)$ with structure constants $c_{ij}^k$, i.e.
\be
\{h_i,h_j\}_{\omega}= \sum_{{k=1}}^{l} c_{ij}^{k}h_{k},\qquad i,j=1,\ldots,l.
 \nonumber
\ee

\item Introduce a Poisson--Hopf algebra deformation $\qlhzd$  of  $C^\infty(\lh^*)$ with  deformation parameter $z\in\mathbb R$ (in a quantum group setting we would have $q:={\rm e}^z$) as the space of smooth functions $F(h_{z,1},\ldots,h_{z,l})$ with fundamental  Poisson bracket given by
\be
\{h_{z,i},h_{z,j}\}_{\omega}= F_{z,ij}(h_{z,1},\dots,h_{z,l }),
\label{zab}
\ee
 where $F_{z,ij}$ are  certain smooth functions also depending smoothly on the deformation parameter $z$ and  such that
\be
\lim_{z\to 0} h_{z,i}=h_i ,\qquad \lim_{z\to 0}\nabla h_{z,i}=\nabla h_i,  \qquad   \lim_{z\to 0}   F_{z,ij}(h_{z,1},\dots,h_{z,l }) =\sum_{k=1}^l c_{ij}^k h_k, 
\label{zac}
\ee
where  $\nabla$ stands for the gradient relative to the Euclidean metric on $\mathbb{R}^2$.
Hence,
\be
  \lim_{z\to 0} \{h_{z,i},h_{z,j}\}_{\omega}=\{h_i,h_j\}_\omega .
\label{zad}
\ee

\item Define the deformed vector fields ${\bf X}_{z,i}$ by the rule
\be
\iota_{{\bf X}_{z,i}}\omega :={\rm d}h_{z,i},
\label{contract2}
\ee
so that
\be
\lim_{z\to 0} {\bf X}_{z,i}= {\bf X}_i.
\label{zae}
\ee

\item Define the   deformed LH system of the initial   system ${\bf X}$  (\ref{aabb}) by
\be
{\bf X}_z:=\sum_{i=1}^lb_i(t){\bf X}_{z,i} .
\label{aabbc}
\ee 
\end{enumerate}

Now some remarks are in order. First, note that for a given LH algebra $\lh$ there exist as many Poisson--Hopf algebra deformations as non-equivalent Lie bialgebra structures $\delta$ on $\lh$~\cite{Chari}, where the 1-cocycle $\delta$ essentially provides the first-order deformation in $z$ of the coproduct map $\Delta$. For three-dimensional real Lie algebras the full classification of Lie bialgebra structures is known~\cite{Gomez}, and some classification results are also known for certain higher-dimensional Lie algebras (see~\cite{Gomez, dualPL, BBM3d} and references therein). Once a specific Lie bialgebra $(\lh,\delta)$ is chosen, the full Poisson--Hopf algebra deformation can be systematically obtained by making use of the Poisson version of the `quantum duality principle'\ for Hopf algebras, as we will explicitly see in the next section for an $(\mathfrak{sl}(2),\delta)$ Lie bialgebra.

Second,  the deformed vector fields ${\bf X}_{z,i}$ (\ref{contract2}) will not, in general, span a finite-dimensional Lie algebra, which implies that  (\ref{aabbc}) is not a Lie system. In fact, the 
sequence of Lie algebra morphisms (\ref{seq}) and the properties of Hamiltonian vector fields \cite{Va94} lead to 
\be
[{\bf X}_{z,i},{\bf X}_{z,j}]= [
\varphi(h_{z,i}),\varphi(h_{z,j})]= \varphi(
\{h_{z,i},h_{z,j}\}_{\omega})= \varphi( F_{z,ij}(h_{z,1},\dots,h_{z,l }))=-\sum_{k=1}^l\frac{\partial F_{z,ij}}{\partial h_{z,k}}{\bf X}_{z,k}.
\nonumber
\ee
In other words,
\be
\left[{\bf X}_{z,i},{\bf X}_{z,j}\right]= \sum_{k=1}^{l} G_{z,ij}^{k} (x,y ) {\bf X}_{z,k},      \label{FRO}
\ee
where the $G_{z,ij}^{k} (x,y)$ are  smooth functions relative to the coordinates $x,y$ and the deformation parameter $z$. 
Despite this,  the relations (\ref{zad}) and the continuity of $\varphi$ imply that
\be 
[{\bf X}_i,{\bf X}_j]=\varphi(\{h_i,h_j\})_\omega=\varphi\left(\lim_{z\rightarrow 0}\{h_{z,i},h_{z,j}\}_\omega\right)=\lim_{z\to 0}\varphi\{h_{z,i},h_{z,j}\}_\omega=\lim_{z\to 0}[{\bf X}_{z,i},{\bf X}_{z,j}] .
\nonumber
\ee
Hence
\be  \lim_{z\to 0}G_{z,ij}^{k} (x,y ) ={\rm constant} 
\nonumber
\ee
holds for all indices.
Geometrically, the conditions (\ref{FRO}) establish that the vector fields ${\bf X}_{z,i}$ span an involutive smooth generalized distribution $\mathcal{D}_z$. In particular, the distribution $\mathcal{D}_0$ is spanned by the Vessiot--Guldberg Lie algebra $\langle {\bf X}_1,\dots,{\bf X}_l\rangle$. This causes $\mathcal{D}_0$ to be integrable on the whole $\mathbb{R}^2$ in the sense of Stefan--Sussman~\cite{Va94,Pa57,WA}. The integrability of $\mathcal{D}_z$, for $z\neq 0$, can only be ensured on open connected subsets of $\mathbb{R}^2$ where $\mathcal{D}_z$ has constant rank~\cite{Va94}.

Third, although the vector fields ${\bf X}_{z,i}$  depend smoothly on $z$, the distribution $\mathcal{D}_z$ may change abruptly. For instance, consider the case given by the LH system ${\bf X}=\partial_x+ty\partial_x$ relative to the symplectic form $\omega=\dd x\wedge \dd y$ and admitting a LH algebra $\lh=\langle h_1:=y,\ h_2:=y^2/2\rangle$. Let us define  $h_{z,1}:=y$ and $h_{z,2}:=y^2/2+zx$. Then ${\bf X}_z=\partial_x+t(y\partial_x-z\partial_y)$ and $\dim \mathcal{D}_0(x,y)=1$, but $\dim \mathcal{D}_z(x,y)=2$ for $z\neq 0$. Hence, the deformation of LH systems may change in an abrupt way the dynamical and geometrical properties of the systems ${\bf X}_z$ (cycles, periodic solutions, etc).
 
Fourth, the deformation parameter $z$  provides an additional degree of freedom that enables the control or modification of the deformed system ${\bf X}_z$. In fact, as $z$ can be taken small, perturbations of the initial Lie system ${\bf X}$ can be obtained from the deformed one ${\bf X}_z$ in a natural way. 

And,  finally, 
we stress that,   by construction,   the very same procedure can be applied to other two-dimensional manifolds different to $\mathbb{R}^2$,  to higher dimensions as well as to multiparameter Poisson--Hopf algebra deformations of Lie algebras endowed with two or more deformation parameters.

\subsect{Constants of the motion}

The fact that $ \qlhzd$ is  a Poisson--Hopf algebra allows us to apply   the coalgebra formalism established in~\cite{BCHLS13Ham}  in order to obtain $t$-independent constants of the motion for ${\bf X}_z$.

Let $S\left(\lh\right)$ be the {\it symmetric algebra} of $\lh$, i.e. the associative unital algebra of polynomials on the elements of $\lh$. The Lie algebra structure on $\lh$ can be extended to a Poisson algebra structure in $S\left(\lh\right)$ by requiring $[v,\cdot ]$ to be a derivation on the second entry for every $v\in \lh$. Then, $S\left(\lh\right)$  can be  endowed with a Hopf algebra structure with a  non-deformed (trivial) coproduct map $\Delta$ defined  by
\begin{equation}
 {\Delta} :S\left(\lh \right)\rightarrow
S\left(\lh\right) \otimes S\left(\lh\right)    ,\qquad      {\Delta}(v_i):=v_i\otimes 1+1\otimes v_i,  \qquad    i=1,\dots, l,
\label{baa}
\end{equation}
which is a Poisson algebra homomorphism relative to the Poisson structure on $S(\lh)$ and the    one induced in $S(\lh)\otimes S(\lh)$. Recall that every element of $S(\lh)$ can be understood as a function on $\lh^*$. Moreover, as  $S(\mathcal{H}_\omega)$ is dense in the space
$C^\infty(\mathcal{H}^*_\omega)$ of smooth functions on the dual $\mathcal{H}^*_\omega$ of the LH algebra $\mathcal{H}_\omega$,  the coproduct in $S(\mathcal{H}_\omega)$ can be extended in a unique way to  
\begin{equation}
 {\Delta} :C^\infty\left(\lh^*\right)\rightarrow
C^\infty\left(\lh^*\right) \otimes C^\infty\left(\lh^*\right).
\nonumber
\end{equation}
Similarly, all structures on $S(\lh)$ can be extended turning  $C^\infty(\mathcal{H}^*_\omega)$ into a Poisson--Hopf algebra. Indeed, the resulting structure is the natural one in $C^\infty(\mathcal{H}^*_\omega)$ given in section 2.2.

Let us assume  now that   $C^\infty\left(\lh^*\right)$  
has a   Casimir  invariant 
\be
C=C(v_1,\dots,v_l),
\nonumber
\ee
where $v_1,\ldots,v_l$ is a basis for $\lh$.   The initial LH system allows us to define a Lie algebra morphism $\phi:\lh\rightarrow C^\infty(M)$, where  $M$ is a submanifold of $\mathbb R^2$ where all functions $h_i:=\phi(v_i)$, for $i=1,\ldots,l$, are well defined.
 Then, the Poisson algebra morphisms 
\be
D: C^\infty\left( \lh^* \right) \rightarrow C^\infty(M),\qquad  D^{(2)} :   C^\infty\left(   \lh^* \right)\otimes C^\infty\left(  \lh^* \right)\rightarrow C^\infty(M)\otimes C^\infty(M),
\label{morphisms}
\ee
defined respectively by
\be
D( v_i):= h_i(x_1,y_1), \qquad
 D^{(2)} \left( {\Delta}(v_i) \right):= h_i(x_1,y_1)+h_i(x_2,y_2)   ,\qquad i=1,\dots, l,
\label{bb}
\ee
lead to the $t$-independent  constants of motion  $F^{(1)}:= F$ and $F^{(2)}$  for the Lie system  ${\bf X}$ given in (\ref{aabb}) where
\be
  F:= D(C),\qquad F^{(2)}:=  D^{(2)} \left( {\Delta}(C) \right).
\label{bc}
\ee

The  very same procedure can also be applied to any   Poisson--Hopf algebra  $ \qlhzd$ with deformed coproduct $\Delta_z$  and Casimir invariant
$C_z=C_z(v_{1},\dots,v_{l})$, where $\{ v_{1},\dots,v_{l} \}$      fulfill the same   Poisson brackets ({\ref{zab}), and  such that
  \be
\lim_{z\to 0} \Delta_{z}=\Delta , \qquad \lim_{z\to 0} C_{z}= C.
\nonumber
\ee
Following~\cite{BCHLS13Ham}, the element $C_z$ turns out to be the cornerstone in the construction of the deformed  constants of the motion  for the `generalized' LH system ${\bf X}_z $.


\sect{A Poisson--Hopf algebra deformation of $\mathfrak{sl}(2)$}

 Once the general description of our  approach  has been  introduced,  we present in this section the general properties of  the Poisson analogue of the so-called non-standard quantum deformation of the simple real Lie algebra $\mathfrak{sl}(2)$. This deformation will be applied in  the sequel   to get deformations of  the    Milne--Pinney equation or Ermakov  system and of some Riccati equations, since all these systems are known to be endowed with a LH algebra $\lh$ isomorphic to  $\mathfrak{sl}(2)$~\cite{BCHLS13Ham, BBHLS,  BHLS}. 

 Let us consider the basis $\{J_3, J_+,J_-\}$ for $\mathfrak{sl}(2)$ with Lie brackets and Casimir operator   given by
\be
[ J_3,J_\pm  ]  = \pm  2J_\pm     ,\qquad  [J_+ , J_- ] = J_3,\qquad {\cal C}=\tfrac 12  J_3^2+(J_+ J_- + J_- J_+).
\label{crules}
\ee
Amongst the three  possible quantum deformations of $\mathfrak{sl}(2)$~\cite{Tmatrix}, we  shall hereafter consider the   non-standard (triangular or Jordanian) quantum deformation, $U_z(\mathfrak{sl}(2))$ (see~\cite{Ohn,beyond, non, Shariati} for further details). The Hopf algebra structure of $U_{z}(\mathfrak{sl}(2))$  has the following 
   deformed   coproduct and  compatible deformed commutation rules  
\be
\Delta_z(J_+)=J_+\otimes 1 + 1 \otimes J_+,\qquad 
\Delta_z(J_j)=J_j\otimes {\rm e}^{2z J_+} + {\rm e}^{- 2z J_+}  \otimes J_j ,\qquad j \in \{-, 3\},
\label{codef}\nonumber
\ee
\be
[ J_3,J_+ ] _z=   \frac{\sinh (  2z J_+)}{  z}    ,\qquad [ J_3, J_-]_z=- J_- \cosh(2zJ_+)  - \cosh(2zJ_+) J_-   ,\qquad  [J_+ , J_- ]_z= J_3.
\label{corudef}\nonumber
\ee
The counit and antipode  can be explicitly found in~\cite{Ohn,non}, and the deformed Casimir reads~\cite{beyond}  
\be
 {\cal C}_z=\frac 12 J_3^2+\frac{\sinh(2z J_+) }{2z} \,  J_- + J_-  \,\frac{\sinh(2z J_+) }{2z} + \frac 12 \cosh^2( 2z J_+) .
\label{bf}\nonumber
\ee

Let $\mathfrak{g}$ be the Lie algebra of $G$.  
 It is well known (see~\cite{Chari,Majid}) that quantum algebras $U_z(\mathfrak{g})$ are Hopf algebra duals of quantum groups $G_z$. On the other hand, quantum groups $G_z$ are just quantizations of Poisson--Lie groups, which are Lie groups endowed with a multiplicative Poisson structure, {i.e.}~a Poisson structure for which the Lie group multiplication is a Poisson map. In the case of $U_z(\mathfrak{sl}(2))$, such Poisson structure on $SL(2)$ is explicitly given by the Sklyanin bracket coming from the classical $r$-matrix
 \be
 r=z J_3\wedge J_+,
 \label{rmns}
 \ee
which is a solution of the (constant) classical Yang--Baxter equation.
 
 Moreover, the `quantum duality principle`~\cite{Dri,STS} states that quantum algebras can be thought of as `quantum dual groups' $G_z^\ast$, which means that any quantum algebra can be obtained as the Hopf algebra quantization of the dual Poisson--Lie group $G^\ast$. The usefulness of this approach to construct explicitly the Poisson analogue of quantum algebras was developed in~\cite{dualPL}.
 
 In the case of $U_z(\mathfrak{sl}(2))$, the Lie algebra $\mathfrak{g}^\ast$ of the dual Lie group $G^\ast$ is given by the dual of the cocommutator map $\delta$ that is obtained from the classical $r$-matrix as
  \begin{equation}
\delta(x)=[ x\otimes 1+1\otimes x , r],\qquad \forall x\in \mathfrak{g}.
\label{rmatrix}
\end{equation}
In our case, from~\eqref{crules} and~\eqref{rmns} we explicitly obtain
\be
\delta(J_3)=2z \, J_3 \wedge J_+ ,\qquad
\delta(J_+)= 0,\qquad
\delta(J_-)= 2z \, J_- \wedge J_+ ,
\nonumber
\ee
and the dual Lie algebra $\mathfrak{g}^\ast$ reads
\begin{equation}
[j^+,j^3]=-2 z \, j^3, \qquad [j^+,j^-]=-2 z \, j^-, \qquad [j^3,j^-]=0,
\label{book}
\end{equation}
where $\{j^3,j^+,j^-\}$ is the basis of $\mathfrak{g}^\ast$,  and $\{J_3,J_+,J_-\}$ can now be  interpreted as local coordinates on the dual Lie group $G^\ast$. The dual Lie algebra~\eqref{book} is the so-called `book' Lie algebra, and the complete set of its Poisson--Lie structures was explicitly obtained in~\cite{BBM3d} (see also~\cite{LV}, where book Poisson--Hopf algebras were used to construct integrable deformations of Lotka--Volterra systems). In particular, if we consider the  coordinates on $G^\ast$ given by
\be
v_1=   J_+,\qquad v_2 =  \tfrac 12 J_3,\qquad v_3= -  J_-,
\nonumber
\ee
the Poisson--Lie structure on the book group whose Hopf algebra quantization gives rise to the quantum algebra $U_z(\mathfrak{sl}(2))$   is given by the fundamental Poisson brackets~\cite{BBM3d}
\be 
\{v_1,v_2\}_z=-\shc (2z v_1)v_1,\qquad 
 \{v_1,v_3\}_z=-2 v_2,\qquad
\{v_2,v_3\}_z= -  \cosh(2 z v_1) v_3,
\label{gb}
\ee
together with the coproduct map
\begin{equation}
 \Delta_z(v_1)=  v_1 \otimes 1+
1\otimes v_1 , \qquad
\Delta_z(v_k)=v_k \otimes \eee^{2 z v_1} + \eee^{-2 z v_1} \otimes
v_k   ,\qquad  k=2,3,
 \label{ga}
 \end{equation}
which is nothing but the group law for the book Lie group $G^\ast$  in the chosen coordinates (see~\cite{dualPL,BBM3d,LV} for a detailed explanation).   Therefore, ~\eqref{gb} and~\eqref{ga}  define a Poisson--Hopf algebra structure on $C^\infty(G^\ast)$, which can be thought of as a Poisson--Hopf algebra deformation of the Poisson algebra $C^\infty(\mathfrak{sl}(2)^\ast)$, since we have identified the local coordinates on $C^\infty(G^\ast)$ with the generators of the Lie--Poisson algebra  $\mathfrak{sl}(2)^\ast$. 

Notice that we have introduced in  (\ref{gb})  the hereafter called  {\it cardinal hyperbolic sinus  function} defined by 
\be
\shc(x):=\frac {\sinh (x)}{x}.
\label{shc}
\ee
Some   properties of  this function along with its relationship with Lie systems are given in the Appendix.

Summarizing, the Poisson--Hopf algebra given by~\eqref{gb} and~\eqref{ga}, together with its Casimir function
 \be
 {C}_z=\shc( 2z v_1)\, v_1v_3-v_2^2   ,
\label{gc}
\ee
will be the deformed Poisson--Hopf algebra that we will use in the sequel in order to construct deformations of LH systems based on $\mathfrak{sl}(2)$. Note that the usual  Poisson--Hopf algebra $C^\infty(\mathfrak{sl}(2)^\ast)$ is smoothly recovered under the $z\to 0$ limit leading to the non-deformed Lie--Poisson coalgebra 
\be
\{ v_1,v_2\}=- v_1,\qquad \{ v_1,v_3\} =- 2v_2,\qquad  \{ v_2,v_3\} =-v_3,
\label{brack2}
\ee 
  with undeformed  coproduct (\ref{baa}) and Casimir
\be
C=v_1 v_3 - v_2^2 .
\label{ai}
\ee

We stress that this application of the `quantum duality principle'  would allow one to obtain the Poisson analogue of any quantum algebra $U_z(\mathfrak{g})$, which by following the method here presented could be further applied in order to construct the corresponding deformation of the LH systems associated to the Lie--Poisson algebra $\mathfrak{g}$. In particular, the Poisson versions of the other quantum algebra deformations of $\mathfrak{sl}(2)$ can be obtained in the same manner with no technical obstructions (for instance, see~\cite{dualPL} for the explicit construction of the `standard' or Drinfel'd--Jimbo deformation).


\sect{Deformed Milne--Pinney  equation and oscillator systems}

As a first application of our approach, we will construct the non-standard deformation of the well-known  Milne--Pinney  (MP) equation \cite{Mi30,Pi50}, which is known to be a LH system~\cite{BBHLS, BHLS}. Recall that the MP equation corresponds  to the equation of motion of the isotropic oscillator with a time-dependent frequency and a `centrifugal' or Rosochatius--Winternitz  term. As we will show in the sequel, the main feature of this deformation is that the new oscillator system has both a position-dependent mass and a time-dependent frequency.


\subsect{Non-deformed  system}

The  MP  equation \cite{Mi30,Pi50} has the following expression
\begin{equation}\label{mp}
\frac{\dd^2x}{\dd t^2}=-\Omega^2(t)x+\frac{c}{x^3},
\end{equation}
where $\Omega(t)$ is any $t$-dependent function and $c\in \mathbb{R}$. 
By introducing a new variable $y:= \dd x/\dd t$, the system  \eqref{mp} becomes a first-order system of differential equations on ${\rm T}\mathbb{R}_0$, where $\mathbb{R}_0:=\mathbb{R}\backslash\{0\}$, of the form
\be
\frac{\dd x}{\dd t}=y,\qquad \frac{\dd y}{\dd t}=-\Omega^2(t)x+\frac{c}{x^3}.
\label{FirstLie}
\ee
This system is indeed part of the one-dimensional Ermakov  system~\cite{Dissertations,Er08,Le91,LA08} and diffeomorphic to the one-dimensional $t$-dependent frequency counterpart~\cite{BCHLS13Ham, BBHLS, BHLS} of the Smorodinsky--Winternitz oscillator \cite{WSUF65}.

The system (\ref{FirstLie})  determines a Lie system  with  associated  $t$-dependent vector field~\cite{BHLS}
\be
{\bf X} ={\bf X}_3+\Omega^2(t){\bf X}_1,
\label{MP}
\ee
where
\begin{equation}\label{FirstLieA}  
{\bf X}_1:=-x\frac{\partial}{\partial y},\qquad {\bf X}_2:=\frac 12 \left(y\frac{\partial}{\partial y}-x\frac{\partial}{\partial x}\right),\qquad {\bf X}_3:=y\frac{\partial}{\partial x}+\frac{c}{x^3}\frac{\partial}{\partial y},
\end{equation}
span a Vessiot--Guldberg Lie algebra $V^{\rm MP}$ of vector fields isomorphic to $\mathfrak{sl}(2)$  (for any value of $c$) with commutation relations given by  
\begin{equation}\label{aa}
[{\bf X}_1,{\bf X}_2]={\bf X}_1,\qquad [{\bf X}_1,{\bf X}_3]=2{\bf X}_2,\qquad [{\bf X}_2,{\bf X}_3]={\bf X}_3 .
\end{equation}
The vector fields of  $V^{\rm MP}$ are defined on $\mathbb R^2_{x\ne 0}$, where they span a regular distribution of order two.

Furthermore,  ${\bf X} $  is a LH system with respect to the symplectic form $\omega={\rm d}x\wedge {\rm d}y$ and the vector fields (\ref{FirstLieA}) admit Hamiltonian functions given by
\be
h_1=\frac 12 x^2 ,\qquad h_2=-\frac 12 xy ,\qquad h_3=\frac 12 \left(y^2 +\frac{c}{x^2} \right),
\label{ham2}
\ee 
that fulfill the following commutation relations with respect  to the Poisson bracket induced by $\omega$:
\be
\{ h_1,h_2\}_\omega=- h_1,\qquad \{ h_1,h_3\}_\omega =- 2h_2,\qquad  \{ h_2,h_3\}_\omega =-h_3 .
\label{brack}
\ee
Then, the functions $h_1,h_2,h_3$ span a LH algebra  ${\cal H}_{\omega}^{\rm {MP}} \simeq \mathfrak{sl}(2)$ of functions on $\mathbb R^2_{x\ne 0}$;  the $t$-dependent Hamiltonian associated with the $t$-dependent vector field (\ref{MP}) reads
\be
h =h_3+\Omega^2(t)h_1   .
\label{hMP}
\ee
We   recall that this Hamiltonian  is a natural one, that is, it can be written in terms of a kinetic energy $T$ and potential $U$
by identifying the variable $y$ as the conjugate momentum $p$  of the coordinate $x$:
\be
h =T+U= \frac 12\, p^2 + \frac 12 \Omega^2(t) x^2 + \frac{c}{2x^2}.
\label{ham}
\ee
 Hence $h$  determines the composition  of a one-dimensional  oscillator with  a time-dependent frequency  $ \Omega(t)$ and unit mass  with a   Rosochatius  or Winternitz potential; the latter  is  just a centrifugal barrier whenever $c>0$ (see~\cite{nonlinear} and references therein).   The LH system  (\ref{FirstLie})  thus comes from the Hamilton equations of $h$ and, obviously, when $c$ vanishes, these reduce  to the equations of motion of a harmonic oscillator with a
time-dependent frequency. 

  We stress that  it has been already proved in~\cite{BBHLS,BHLS} that the MP equations (\ref{FirstLie}) comprise the {\em three} different types of possible $\mathfrak{sl}(2)$-LH systems according to the value of the constant $c$:   class  {\rm P}$_2$ for $c>0$;  class {\rm I}$_4$ for $c<0$; and class {\rm I}$_5$ for $c=0$. This means that any other LH system related to a Vessiot--Guldberg Lie algebra of Hamiltonian vector fields isomorphic to $\mathfrak{sl}(2)$ must be, up to a $t$-independent change of variables, of the form (\ref{FirstLie}) for a positive, zero or negative value of $c$.
This implies that the second-order Kummer--Schwarz equations~\cite{CGL11, LS12} and several  types of  Riccati equations~\cite{Mariton, Eg07, Wi08, SSVG11, SSVG14,CGLS, pilar} are comprised within  ${\cal H}_{\omega}^{\rm {MP}}$ (depending on the sign of $c$). The relationships amongst all of these systems are  ensured by construction and these can be explicitly obtained  through either diffeomorphisms or changes of variables (see~\cite{BBHLS,BHLS} for details).

The constants of motion for the MP equations can be obtained by applying the coalgebra formalism introduced in~\cite{BCHLS13Ham} and briefly summarized in section 2.4. Explicitly, 
  let us consider the   Poisson--Hopf algebra $C^\infty({\cal H}_{\omega}^{\rm {MP}*} )$   with basis  $\{ v_1,v_2,v_3\}$,  coproduct    (\ref{baa}), fundamental Poisson brackets  (\ref{brack2}) and Casimir (\ref{ai}).
   The Poisson algebra morphisms  (\ref{morphisms}) 
 \be
 D: C^\infty({\cal H}_{\omega}^{\rm {MP}*}) \rightarrow C^\infty(\mathbb R^2_{x\ne 0}) ,\quad D^{(2)} :    C^\infty( {\cal H}_{\omega}^{\rm {MP}*} ) \otimes  C^\infty ( {\cal H}_{\omega}^{ \rm {MP}*} )\rightarrow C^\infty(\mathbb R^2_{x\ne 0})\otimes C^\infty(\mathbb R^2_{x\ne 0}) ,
 \nonumber
 \ee
  defined by (\ref{bb}),  where $h_i$ are the Hamiltonian functions   (\ref{ham2}), lead to the   $t$-independent  constants of the motion  $F^{(1)}:=F$ and $F^{(2)}$  given by (\ref{bc}), through   the Casimir (\ref{ai}),   for the Lie system  ${\bf X}$ (\ref{FirstLie}); namely~\cite{BCHLS13Ham}
\bea
&& F= h_1(x_1,y_1) h_3(x_1,y_1)- h_2^2(x_1,y_1)=\frac c 4 ,\nonumber\\[2pt]
&&F^{(2)}=\bigl( \left[ h_1(x_1,y_1)+h_1(x_2,y_2)\right] \left[ h_3(x_1,y_1)+h_3(x_2,y_2)\right]  \bigr) -\bigl( h_2(x_1,y_1)+h_2(x_2,y_2) \bigl)^2\nonumber\\[2pt]
&&\qquad\, =  \frac 14 ({x_1}{y_2} -{x_2} {y_1})^2 
+\frac c 4\,  \frac{(x_1^2+x_2^2)^2}{x_1^2 x_2^2} .
\label{am}
\eea
We observe that $F^{(2)}$ is just a Ray--Reid invariant for generalized Ermakov systems \cite{Le91,RR79} and that  it is related to  the one obtained in~\cite{coalgebra2,letterBH} from a coalgebra approach  applied to superintegrable systems. 

 By permutation of the indices corresponding to the variables of  the non-trivial invariant $F^{(2)}$,  we find two other constants of the motion:
\be 
    F_{13}^{(2)}=S_{13} ( F^{(2)}   ) ,\qquad    F_{23}^{(2)}=S_{23} ( F^{(2)}   ) ,
\label{an}
\ee
 where $S_{ij}$ is the permutation of variables $(x_{i},y_i)\leftrightarrow
(x_j,y_j)$. Since    $\partial(F^{(2)},F^{(2)}_{23})/\partial(x_1,y_1)\neq 0$, both constants of motion are functionally independent (note that the pair $(F^{(2)},F^{(2)}_{13})$  is functionally independent as well). From these two invariants, the corresponding superposition rule can be  derived in a straightforward manner. Its explicit expression can be found in~\cite{BCHLS13Ham}.


\subsect{Deformed Milne--Pinney  equation}

In order to apply the non-standard deformation of $\mathfrak{sl}(2)$ described in section 3  to the MP equation, we need to find  the deformed counterpart $h_{z,i}$ $(i=1,2,3)$ of the Hamiltonian functions $h_i$ (\ref{ham2}), so fulfilling the   Poisson brackets  (\ref{gb}),   by  keeping the canonical symplectic form $\omega$. 

 This problem can be rephrased as the one consistent in finding symplectic realizations of a given Poisson algebra, which can be solved once a particular symplectic leave is fixed as a level set for the Casimir functions of the algebra, where the generators of the algebra can be  expressed in terms of the corresponding Darboux coordinates. 
In the particular case of the $U_z(\mathfrak{sl}(2))$ algebra, the explicit solution (modulo canonical transformations) was obtained in~\cite{chains} where
the algebra~\eqref{gb} was found to be generated by the functions
\bea 
&&v_1(q,p)=\frac 12\,q^2,\cr
&&v_2(q,p)=-\frac 12\frac {\sinh z q^2}{z q^2} \, q p , \cr
&&v_3(q,p)=\frac 12\frac {\sinh z q^2}{z q^2}\,  p^2 +
\frac 12\frac{z c}{\sinh z q^2},
\nonumber
\eea
where $\omega={\rm d} q\wedge {\rm d} p$, and the Casimir function~\eqref{gc} reads ${C}_z=c/4$. In practical terms, such a solution can easily be  found by solving firstly the non-deformed case $z\to 0$ and, afterwards, by deforming the $v_i(q,p)$ functions under the constraint that the Casimir ${C}_z$ has to take a constant value.
With this result at hand, the corresponding deformed vector fields ${\bf X}_{z,i}$ can   be computed by imposing the relationship (\ref{contract2}) and the final result is summarized in the following statement. 
  
\begin{proposition}
\label{proposition1} (i) The Hamiltonian functions defined by
\be
h_{z,1}:=\frac 12 x^2 ,  \qquad
 h_{z,2}:= -      \frac 12\shc (z x^2)\, x  y    , \qquad
 h_{z,3}:= \frac 12 \left(\! \shc (z x^2) \, y^2 + \frac 1{ \shc (zx^2)}\,
\frac{c}{x^2}  \right)  ,
\label{gd}
\ee
 close the Poisson brackets (\ref{gb}) with respect to the symplectic form  $\omega={\rm d}x\wedge {\rm d}y$ on $\mathbb R^2_{x\ne 0}$, namely
\begin{equation}\label{gb2}
\begin{gathered}
\{h_{z,1},h_{z,2}\}_\omega=-\shc (2z h_{z,1} )h_{z,1},\qquad 
 \{h_{z,1},h_{z,3}\}_\omega=-2 h_{z,2},\\[2pt]
\{h_{z,2},h_{z,3}\}_\omega= - \cosh(2 z h_{z,1})  h_{z,3},
\end{gathered}
\end{equation}
where $\shc(x)$ is defined in (\ref{shc}).
 Relations~\eqref{gb2} define  the deformed Poisson  algebra $C^\infty({\cal H}_{z,\omega}^{\rm {MP}*})$. 

\noindent
 (ii)  The vector fields ${\bf X}_{z,i}$ corresponding to $h_{z,i}$  read
\begin{equation}
\begin{gathered}
{\bf X}_{z,1}   =   -x\frac{\partial}{\partial y},\qquad {\bf X}_{z,2}=\left(\cosh(zx^{2})-\frac 12 \shc(zx^{2})\right)y\frac{\partial}{\partial y}-\frac 12 \shc(zx^{2})\,x\frac{\partial}{\partial x}, \\
{\bf X}_{z,3}   =   \shc(zx^{2})\,y\frac{\partial}{\partial x}+\left[\frac{c}{x^{3}}\, \frac{\cosh(zx^{2})}{\shc^{2}(zx^{2})} +\frac{\shc(zx^{2})-\cosh(zx^{2})}{x}\,y^{2}\right]\frac{\partial}{\partial y},\nonumber
\end{gathered}
\end{equation}
which satisfy 
\begin{equation}
\begin{gathered}
\left[{\bf X}_{z,1},{\bf X}_{z,2}\right]=\cosh (z x^2) \, {\bf X}_{z,1},\qquad [{\bf X}_{z,1},{\bf X}_{z,3}]=2 {\bf X}_{z,2}, \\[2pt] 
[{\bf X}_{z,2},{\bf X}_{z,3}]=\cosh (z x^2) \, {\bf X}_{z,3}+ z^2\left(    c  + x^2  y^2\,  \shc^2 (z x^2)  \right)  {\bf X}_{z,1}. 
\label{com2}
\end{gathered}
\end{equation}
\end{proposition}

\smallskip

Since $\lim_{z\to 0}\shc(z x^2)=1$ and $\lim_{z\to 0}\cosh(z x^2)=1$, it can directly  be checked that
 all the classical limits   (\ref{zac}), (\ref{zad}) and (\ref{zae}) are fulfilled. As expected,  the Lie derivative of $\omega$  with respect to each ${\bf X}_{z,i}$  vanishes.

At this stage, it is important to realize that, albeit (\ref{gb2}) are genuine  Poisson brackets defining the Poisson algebra   $C^\infty({\cal H}_{z,\omega}^{\rm {MP}*})$, the commutators (\ref{com2}) show that ${\bf X}_{z,i}$ do not span a new Vessiot--Guldberg Lie algebra; in fact, the commutators give rise to linear combinations of the vector fields ${\bf X}_{z,i}$ with coefficients that are functions depending on the coordinates and the deformation parameter.

  Consequently, proposition~\ref{proposition1} leads to  a deformation of the initial Lie system (\ref{MP}) and of the LH one (\ref{hMP})  defined by
   \be
{\bf X}_z:={\bf  X}_{z,3}+\Omega^2(t){\bf X}_{z,1},\qquad h_z:=h_{z,3}+\Omega^2(t)h_{z,1}.
\label{MPz}
\ee
Thus we    obtain  the following $z$-parametric   system of differential equations that generalizes (\ref{FirstLie}):
\bea
&&\frac{\dd x}{\dd t}=\shc (z x^2)\, y,\nonumber\\[2pt]
&& \frac{\dd y}{\dd t}=-\Omega^2(t)x+    
\frac{c}{x^3} \, \frac{\cosh (z x^2)  }{ \shc^2(z x^2) }+ \frac{\shc (z x^2)- \cosh (z x^2)}{x}  \, y^2 .
\label{FirstLie2}
\eea
From the first equation, we can write   $$y=\frac 1{   \shc (z x^2) }\, \frac{\dd x }{\dd t},
$$
and by substituting this expression into the second equation in (\ref{FirstLie2}), we obtain   
a deformation of the MP equation~\eqref{mp}  in the form
\be
\frac{\dd^2 x}{\dd t^2}  + \left(\frac{1}{x}- \frac{z x}{ \tanh (z x^2)} \right)   \biggl(\frac{\dd x}{\dd t} \biggr)^2 =-\Omega^2(t) \, x  \shc (z x^2)+ \,\frac{c\,z }{x \tanh (z x^2)} .
\nonumber
\ee
Note that this really is    a deformation of the MP equation in the sense that   the limit $z\to 0$   recovers  the standard one (\ref{mp}).


\subsect{Constants of motion for the deformed  Milne--Pinney system}

 An essential feature of the formalism here presented is the fact that  $t$-independent constants of motion for the deformed system ${\bf X}_z$ (\ref{MPz}) can be deduced by using the coalgebra structure of $C^\infty( {\cal H}_{z,\omega}^{\rm {MP}*})$.  Thus we start with  the Poisson--Hopf algebra    $C^\infty( {\cal H}_{z,\omega}^{\rm {MP}*} )$ with deformed coproduct $\Delta_z$ given by (\ref{ga}) and, following section 2.4~\cite{BCHLS13Ham}, we consider  the Poisson algebra morphisms 
\be
D_z: C^\infty( {\cal H}_{z,\omega}^{\rm {MP}*} )\rightarrow C^\infty( \mathbb R^2_{x\ne 0}),\quad  D_z^{(2)} : C^\infty( {\cal H}_{z,\omega}^{\rm {MP}*} )\otimes   C^\infty( {\cal H}_{z,\omega}^{\rm {MP}*} )\rightarrow C^\infty(\mathbb R^2_{x\ne 0})\otimes C^\infty(\mathbb R^2_{x\ne 0}),
\nonumber
\ee
which are  defined by   
\bea 
&& D_z( v_i)= h_{z,i}(x_1,y_1):=  h_{z,i}^{(1)}  , \quad i=1,2,3, \nonumber\\ 
&&  D_z^{(2)} \left( {\Delta}_z(v_1) \right) = h_{z,1}(x_1,y_1)+h_{z,1}(x_2,y_2):= h_{z,1}^{(2)} \,  ,  \nonumber \\
&& 
D_z^{(2)} \left( {\Delta}_z(v_j) \right) = h_{z,j}(x_1,y_1)  {\rm e}^{2 z h_{z,1}(x_2,y_2)}  + {\rm e}^{-2 z h_{z,1}(x_1,y_1)} h_{z,j}(x_2,y_2):=  h_{z,j}^{(2)} \,  ,\quad j= 2,3,
\nonumber
\eea
where $h_{z,i}$ are the Hamiltonian functions (\ref{gd}), so fulfilling (\ref{gb2}). Hence (see \cite{chains})
\bea
&&  h_{z,1}^{(2)} =  \frac 12(x_1^2+x_2^2)  ,\nonumber\\[2pt]
&&  h_{z,2}^{(2)} =  -      \frac 12 \left(  \! {\shc (z x_1^2)}  \, x_1  y_1  {\rm e}^{z x_2^2}   +   {\rm e}^{- z x_1^2}      {\shc (z x_2^2)}  \, x_2  y_2 \right)    , \nonumber\\[2pt]
&&  h_{z,3}^{(2)} =\frac  12   \left(  \!  {\shc (z x_1^2)}\,  y_1^2 +
\frac{c}{ x_1^2 \shc (zx_1^2)} \right) {\rm e}^{z x_2^2}   + \frac 12\,   {\rm e}^{- z x_1^2}   \left(  \! {\shc (z x_2^2)} \, y_2^2 +
\frac{c}{ x_2^2\, \shc (zx_2^2)} \right). 
\nonumber
\eea
Recall that, by construction,  the functions $h_{z,i}^{(2)}$   fulfill the   Poisson brackets (\ref{gb2}). 
 The
$t$-independent  constants of motion are  then obtained through 
$$
  F_z= D_z(C_z),\qquad F_z^{(2)}=  D_z^{(2)} \left( {\Delta_z}(C_z) \right),
$$
where  $C_z$ is   the Casimir (\ref{gc}); these are 
\bea
&& \!\!\!\!  \!\!\!\!  \!\!\!\! \!  F_z=  \shc\bigl(2 z h_{z,1}^{(1)}  \bigr)  h_{z,1}^{(1)}   h_{z,3}^{(1)} - \bigl( h_{z,2}^{(1)}  \bigr)^2 = \frac c    4 \,  ,\nonumber\\[2pt]
&& \!\!\!\!  \!\!\!\!  \!\!\!\! \!  F_z^{(2)}=  \shc\bigl(2 z h_{z,1}^{(2)} \bigr)  h_{z,1}^{(2)}   h_{z,3}^{(2)} - \bigl( h_{z,2}^{(2)}  \bigr)^2  \label{amz}\\[2pt]
&& = 
 \frac 14 \left[  \shc(z x_1^2)      \shc(z x_2^2) \,  ({x_1}{y_2} -{x_2} {y_1})^2     +c\,  \frac{ \shc^2\bigl( z(x_1^2+x_2^2) \bigr)}{\shc( z x_1^2) \shc( z x_2^2)}\,  \frac{(x_1^2+x_2^2)^2}{x_1^2 x_2^2} \right]  {\rm e}^{- z x_1^2} {\rm e}^{z x_2^2} ,
\nonumber
\eea
so providing the corresponding deformed  Ray--Reid invariant, being (\ref{am}) its non-deformed counterpart  with $z=0$. Notice that this invariant is related  to the so-called `universal   constant of the motion'  coming from  $U_z(\mathfrak{sl}(2))$  and given in~\cite{letterBH}.
As in (\ref{an}), other  equivalent constants of motion can be deduced from $ F_z^{(2)}$ by permutation of the variables.


\subsect{A new  oscillator system with position-dependent mass}

 If we set $p:= y$, the $t$-dependent Hamiltonian $h_z$  in (\ref{MPz})  can be written, through (\ref{gd}),  as:
\be
h_z= T_z+U_z= \frac 12  {\shc (z x^2)}\,  p^2+ \frac 12\Omega^2(t) x^2  +
\frac{ c}{2 x^2  {\shc (z x^2)} }  \,  , 
\nonumber
\ee
so deforming $h$ given in  (\ref{ham}). The corresponding Hamilton equations are just   (\ref{FirstLie2}). 

  It is worth mentioning that $h_z$ can be interpreted naturally within the framework of 
 position-dependent mass oscillators (see~\cite{CrNN07, Quesne07, CrR09, BurgosAnnPh11, Ran14Jmp, GhoshRoy15, MustJpa15, Quesne15Jmp} and references therein). The above Hamiltonian naturally suggests the definition of a position-dependent mass function in the form
\be
m_z(x) :=\frac 1{  \shc (z x^2) }= \frac{z x^2}{ \sinh (z x^2)}  \, ,\qquad \lim_{z\to 0} m_z(x) =1,  \qquad \lim_{x\to \pm\infty } m_z(x) =0.
\label{masa}
\ee
Then $h_z$ can be rewritten as
\be
h_z=  \frac {p^2}{2m_z(x)}  + \frac 12m_z(x) \Omega^2(t)\left[ x^2  \shc (z x^2)   \right]  +
\frac{  c}{2m_z(x)} \left[  \frac{1}{x ^2\shc^2(z x^2)} \right]   .
\nonumber
\ee
Thus  the Hamiltonian $h_z$ can be regarded as a system corresponding to   a particle with  position-dependent mass $m_z(x)$ under a deformed oscillator potential  $U_{z,{\rm osc}}(x)$ with time-dependent frequency  $\Omega(t)$ and   a deformed  Rosochatius--Winternitz potential $U_{z,{\rm RW}}(x)$ given by
\bea
&& 
U_{z,{\rm osc}}(x): =  x^2 \shc (z x^2)=  \frac{\sinh (z x^2)}{z} \, ,\label{oscz}\\
&& 
 U_{z,{\rm RW}}(x):=\frac{1}{x ^2\shc^2(z x^2)}= \left( \frac{z x}{\sinh (z x^2)}\right)^2,
\nonumber
\eea
such that
\bea
&&  \lim_{z\to 0} U_{z,{\rm osc}}(x) =x^2,  \qquad\  \lim_{x\to \pm\infty }U_{z,{\rm osc}}(x)  =+\infty, \nonumber\\[2pt]
&&  \lim_{z\to 0}  U_{z,{\rm RW}}(x)=\frac 1{x^2},  \qquad\  \lim_{x\to \pm\infty } U_{z,{\rm RW}}(x) =0 .
\nonumber 
\eea
The deformed mass and the oscillator potential functions are represented in figures 1 and  2.}

The Hamilton equations (\ref{FirstLie2}) can easily be  expressed in terms of $m_z(x)$ as
\bea
&&\!\!\!\!\!\!\!\!\!\!   \dot x = \frac{\partial h^{\rm MP}_z}{\partial p}= \frac p{m_z(x)} ,\nonumber\\
&&\!\!\!\!\!\!\!\!\!\!   \dot p= -  \frac{\partial h^{\rm MP}_z}{\partial x} = -m_z(x)  \Omega^2(t)  \, x \shc (z x^2) + \frac{c}{m_z(x)} \, \frac {\cosh (z x^2)} {x^3 \shc^3 (z x^2)}   +p^2  \frac{ m^\prime_z(x)}{2 m^2_z(x)},
\nonumber
\eea
and the constant of the motion (\ref{amz})   turns out to be
\be
 F_z^{(2)}=    \frac 14 \left[  \frac {({x_1}{p_2} -{x_2} {p_1})^2 }{m_z(x_1)m_z(x_2) }    +c\, m_z(x_1)m_z(x_2)    { \shc^2\bigl( z(x_1^2+x_2^2) \bigr)} \,  \frac{(x_1^2+x_2^2)^2}{x_1^2 x_2^2} \right]  {\rm e}^{- z x_1^2} {\rm e}^{z x_2^2}.
\nonumber
\ee


\begin{figure}[t]
\begin{center}
\includegraphics[height=6.0cm]{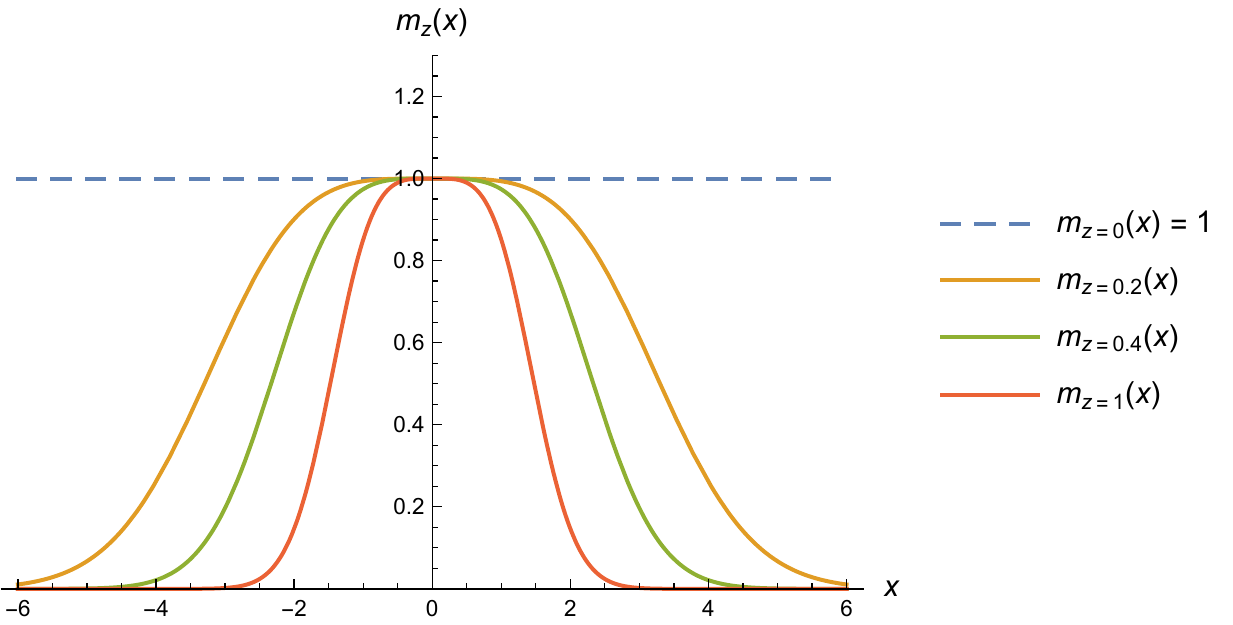}
\caption{\small The position-dependent mass (\ref{masa}) for different values of the deformation parameter $z$.}
 \label{pdm}
\end{center}
\end{figure}



\begin{figure}[t]
\begin{center}
\includegraphics[height=6.0cm]{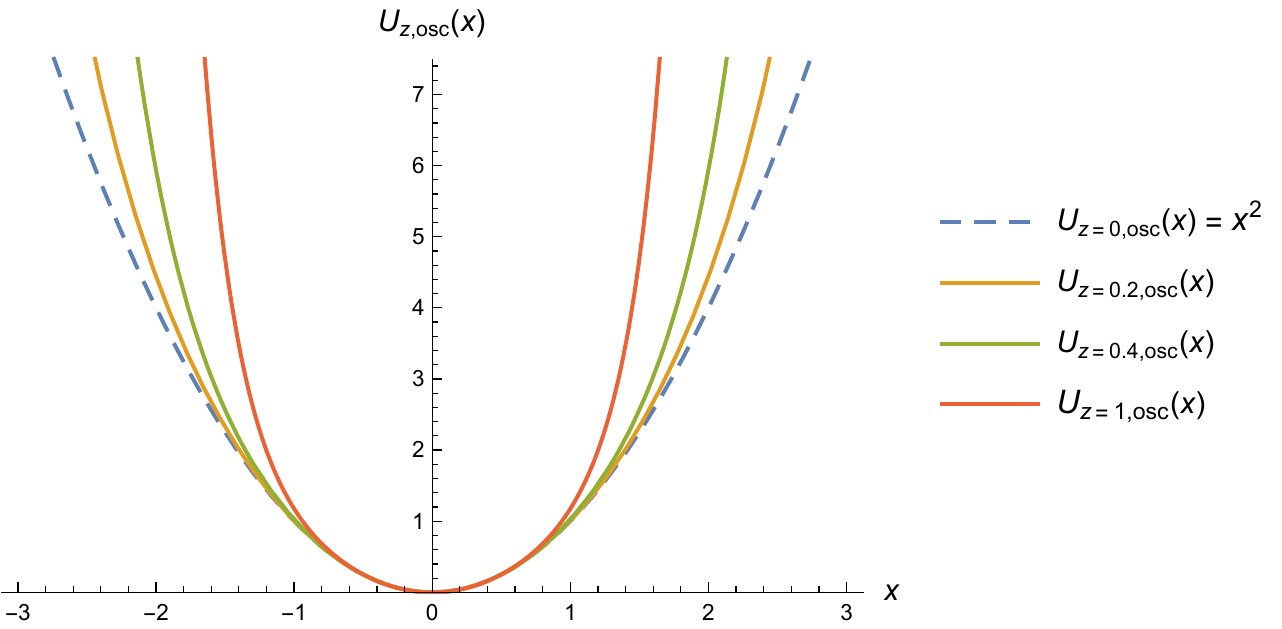}
\caption{\small The deformed oscillator potential  (\ref{oscz}) for different values of the deformation parameter $z$.}
 \label{osc}
\end{center}
\end{figure}



\sect{Deformed complex Riccati  equation}

 In this section we  consider the   complex Riccati equation given by
\begin{equation}
\frac{{\rm d} z}{{\rm d} t}=b_1(t)+b_2(t)z+b_3(t)z^2,\qquad z\in\mathbb{C},
\label{da}
\end{equation}
where $b_i(t)$  are arbitrary $t$-dependent real coefficients. We recall that (\ref{da}) is related to certain planar Riccati equations~\cite{Eg07,Wi08} and that several mathematical and physical applications can be found in~\cite{Ju97,FMR10,Or12,Sc12}.

By writing $z= \xx+i \yy$, we find that (\ref{da}) gives rise to a system of the type (\ref{system}), namely 
\begin{equation}
\frac{{\rm d} \xx}{{\rm d} t}=b_1(t)+b_2(t)\xx+b_3(t)(\xx^2- \yy^2),\qquad \frac{{\rm d} \yy}{{\rm d} t}=b_2(t)\yy+2b_3(t)\xx \yy .
\label{db}
\end{equation}
Thus the  associated  $t$-dependent vector field  reads
\be
 {\bf X}=b_1(t){\bf X}_1+b_2(t){\bf X}_2+b_3(t){\bf X}_3,
 \label{dc}
\ee
 where
\begin{equation}
{\bf X}_1= \frac{\partial}{\partial \xx},\qquad {\bf X}_2= \xx\frac{\partial}{\partial \xx}+\yy\frac{\partial}{\partial \yy} ,\qquad {\bf X}_3= (\xx^2- \yy^2)\frac{\partial}{\partial \xx}+2\xx \yy\frac{\partial}{\partial \yy} ,
\label{vectRiccati2}
\end{equation}
span a Vessiot--Guldberg Lie algebra $V^{\rm CR}\simeq \mathfrak{sl}(2)$ with the same commutation relations (\ref{aa}). It has already be proven that the system ${\bf X}$  is a LH one belonging to the  class  {\rm P}$_2$~\cite{ BBHLS, BHLS} and  that their vector fields span a regular distribution on $\mathbb{R}^2_{\yy\neq 0}$. The  symplectic form, coming from (\ref{der}),  and    the corresponding Hamiltonian functions (\ref{contract})   turn out to be
 \be
\omega=\frac{\dd \xx\wedge \dd \yy}{\yy^2},\qquad h_1= -\frac 1{\yy},\qquad h_2= -\frac \xx \yy , \qquad h_3=- \frac{\xx^2+\yy^2}{\yy} ,
\label{de}
\ee
 which fulfill the commutation rules (\ref{brack}) so defining   a LH algebra  ${\cal H}_{\omega}^{\rm {CR}}$.  
A $t$-dependent Hamiltonian associated with ${\bf X}$ reads
\be
h=b_1(t)h_1+b_2(t)h_2+b_3(t)h_3   .
\label{de2}
\ee
In this case, the constants of the motion     (\ref{bc}) are found to be $F=1$ and~\cite{BHLS}
\be
 F^{(2)}= \frac{(\xx_1-\xx_2)^2+(\yy_1+ \yy_2)^2}{\yy_1 \yy_2} \, .
 \label{df}
 \ee

As commented above,  the  Riccati system (\ref{db}) is locally diffeomorphic  to the MP equations (\ref{FirstLie})  with  $c>0$, both belonging to the same   class  {\rm P}$_2$~\cite{BBHLS}. Explicitly,  the change of variables
\be
x=\pm \frac{c^{1/4}}{ \sqrt{|v| }},\qquad  y=\mp \frac{c^{1/4} \, u}{ \sqrt{|v|}},\qquad u=-\frac{y}x,\qquad |v|=  \frac{c^{1/2}}{x^2},\qquad c>0,
\label{dg}
\ee
map, in this order,  the vector fields (\ref{FirstLieA}) on $\mathbb{R}^2_{x\neq 0}$, the symplectic form $\omega={\rm d}x\wedge {\rm d}y$, Hamiltonian functions (\ref{ham2}) and the constant of motion (\ref{am}) 
onto the vector fields (\ref{vectRiccati2}) on $\mathbb{R}^2_{\yy\neq 0}$, (\ref{de}) and (\ref{df}) (up to a multiplicative constant $\pm\frac 12 c^{1/2}$).

To obtain the corresponding   (non-standard) deformation of the complex Riccati  system (\ref{db}),   the very same change of variables (\ref{dg}) can be considered since, in our approach, the symplectic form (\ref{de}) is kept non-deformed.  Thus, by starting from proposition~\ref{proposition1}  and applying (\ref{dg}) (with $c=4$ for simplicity), we get the following result.

\begin{proposition}
\label{proposition2}
(i) The Hamiltonian functions given by
\be
h_{z,1}=- \frac 1 {\yy} \, ,  \qquad
 h_{z,2}= - \shc (2z /\yy)  \,    \frac \xx\yy  \,  , \qquad
 h_{z,3}=-  \frac{ \shc^2 (2z /\yy) \, \xx^2 + \yy^2 }{\shc (2z /\yy)\,  \yy}    \, ,
\nonumber
\ee
fulfill the  commutation rules (\ref{gb2}) with respect to the Poisson bracket induced by the symplectic form  $\omega$ (\ref{de})   defining the deformed Poisson algebra $C^\infty({\cal H}_{z,\omega}^{\rm {CR}*})$. 

\noindent
(ii) The   corresponding   vector fields ${\bf X}_{z,i}$ read
\begin{eqnarray}
&& {\bf X}_{z,1}   =  \frac{\partial}{\partial \xx},\qquad {\bf X}_{z,2}=\xx \cosh(2z/ \yy)\frac{\partial}{\partial \xx} + \yy  \shc(2z/  \yy) \frac{\partial}{\partial \yy} , \nonumber\\
&& {\bf X}_{z,3}   =    \left(\xx^2 -\frac{\yy^2}{ \shc^2(2z/ \yy)} \right)   \cosh(2z /\yy)\frac{\partial}{\partial \xx}+2\xx\yy  \shc(2z/ \yy)\frac{\partial}{\partial \yy} ,\nonumber
\end{eqnarray}
which satisfy 
\bea
&& [{\bf X}_{z,1},{\bf X}_{z,2}]=\cosh (2z/ \yy) \, {\bf X}_{z,1},\qquad [{\bf X}_{z,1},{\bf X}_{z,3}]=2 {\bf X}_{z,2},\nonumber\\[2pt] 
&& [{\bf X}_{z,2},{\bf X}_{z,3}]=\cosh (2z/ \yy) \, {\bf X}_{z,3}+ 4 z^2\left( 1  + \frac{\xx^2}{\yy^2}  \shc^2 (2z/ \yy)  \right)  {\bf X}_{z,1}. 
\nonumber
\eea
\end{proposition}

Next the deformed counterpart of the Riccati Lie system (\ref{dc}) and of the   LH one (\ref{de2}) is defined by
 \be
{\bf X}_z:=b_1(t){\bf X}_{z,1}+b_2(t){\bf X}_{z,2}+b_3(t){\bf X}_{z,3},\qquad h_z:=b_1(t)h_{z,1}+b_2(t)h_{z,2}+b_3(t)h_{z,3} .
\label{dl}
\ee
And the   $t$-independent constants of motion turn out to be $F_z=1$ and
\be
 F_z^{(2)}=\left(   \shc (2z/ \yy_1)   \shc (2z/ \yy_2)   \frac{(\xx_1-\xx_2)^2 }{\yy_1 \yy_2} + \frac{\shc^2 (2z/ \yy_1+ 2z/ \yy_2) }{  \shc (2z/ \yy_1)   \shc (2z/ \yy_2)  } \frac{ (\yy_1+ \yy_2)^2}{\yy_1 \yy_2}  \right) \eee^{2z/\yy_1}  \eee^{-2z/\yy_2} \, .
\nonumber
\ee
Therefore  the   deformation of the system (\ref{db}), defined by ${\bf X}_z$ (\ref{dl}),  reads
\bea
&& \frac{{\rm d} \xx}{{\rm d} t}=b_1(t)+b_2(t) \xx \cosh(2z/ \yy)  +b_3(t)  \left(\xx^2 -\frac{\yy^2}{ \shc^2(2z/ \yy)} \right)   \cosh(2z /\yy), \nonumber\\[2pt]
&& \frac{{\rm d} \yy}{{\rm d} t}=b_2(t)  \yy  \shc(2z/  \yy)+2b_3(t) \xx\yy  \shc(2z/ \yy) .
\nonumber
\eea


\sect{Deformed coupled Riccati  equations}

As a last application, let us consider two coupled Riccati equations given by
 \cite{Mariton}
\begin{equation}
\frac{{\rm d}\xx}{{\rm d}t}=a_0(t)+a_1(t)\xx+a_2(t)\xx^2,\qquad \frac{{\rm d}\yy}{{\rm d}t}=a_0(t)+a_1(t)\yy+a_2(t)\yy^2,
\label{ea}
\end{equation}
constituting a particular case of the systems of Riccati equations studied in~\cite{BCHLS13Ham,CGLS}.  

Clearly, the system (\ref{ea}) is a Lie system associated with a $t$-dependent vector field
\be
 {\bf X}=a_0(t){\bf X}_1+a_1(t){\bf X}_2+a_2(t){\bf X}_3,
 \label{eb}
\ee
where
\begin{equation}
{\bf X}_1= \frac{\partial}{\partial \xx}+ \frac{\partial}{\partial \yy},  \qquad {\bf X}_2= \xx\frac{\partial}{\partial \xx}+\yy\frac{\partial}{\partial \yy} ,\qquad {\bf X}_3= \xx^2\frac{\partial}{\partial \xx}+\yy^2\frac{\partial}{\partial \yy} ,
\label{ec}
\end{equation}
close on the commutation rules (\ref{aa}), so  spanning  a Vessiot--Guldberg Lie algebra $V^{\rm 2R}\simeq \mathfrak{sl}(2)$. 
Furthermore, ${\bf X}$  is a LH system which belongs to the class  {\rm I}$_4$~\cite{ BBHLS, BHLS} restricted to $\mathbb{R}^2_{\xx\neq \yy}$.
The symplectic form and Hamiltonian functions for ${\bf X}_1,{\bf X}_2, {\bf X}_3$ read   
 \be
\omega=\frac{\dd \xx\wedge \dd \yy}{(\xx-\yy)^2} ,\qquad h_1= \frac 1{\xx-\yy},\qquad h_2= \frac 12\left( \frac {\xx+\yy}  {\xx-\yy}\right) , \qquad h_3= \frac{\xx \yy}{\xx-\yy}.
\label{ee}
\ee
The functions $h_1,h_2,h_3$ satisfy the   commutation rules (\ref{brack}),  thus spanning a LH algebra  ${\cal H}_{\omega}^{\rm {2R}}$.  
Hence, the $t$-dependent Hamiltonian associated with ${\bf X}$  is given by
\be
h=a_0(t)h_1+a_1(t)h_2+a_2(t)h_3  .
\label{ee2}
\ee
The   constants of the motion     (\ref{bc}) are  now $F=-1/4$ and~\cite{BHLS}
\be
 F^{(2)}=-  \frac{ (\xx_2- \yy_1  ) (\xx_1- \yy_2  )}    { (\xx_1- \yy_1  ) (\xx_2- \yy_2  )} \, .
 \label{ef}
 \ee

The LH system    (\ref{ea}) is locally diffeomorphic  to the MP equations (\ref{FirstLie}) but now  with  $c<0$~\cite{BBHLS}. 
Such a diffeomorphism is achieved through the     change of variables given by
\bea
&& x=\pm \frac{ (4|c|)^{1/4}}{ \sqrt{|\xx-\yy| }},\qquad  y=\mp \frac { (4|c|)^{1/4} (\xx+\yy)}{2 \sqrt{|\xx-\yy| }},\qquad  c<0, \nonumber\\[2pt]
&& \xx=\pm \frac{ |c|^{1/2}}{x^2}-\frac y x ,\qquad \yy=\mp \frac{ |c|^{1/2}}{x^2}-\frac y x  ,
\label{eg}
\eea
which map  the MP vector fields (\ref{FirstLieA}) with domain   $\mathbb{R}^2_{x\neq 0}$, symplectic form $\omega={\rm d}x\wedge {\rm d}y$, Hamiltonian functions (\ref{ham2}) and constant of motion (\ref{am}) 
  onto (\ref{ec}) with domain $\mathbb{R}^2_{\xx\neq \yy}$, (\ref{ee}) and (\ref{ef}) (up to a multiplicative constant $\pm |c|^{1/2}$), respectively.

As in the previous section, the   (non-standard) deformation of the coupled Riccati    system (\ref{ea})  is obtained by   starting again from proposition~\ref{proposition1}  and now    applying the change of variables (\ref{eg}) with $c=-1$ (without loss of generality) finding the following result.

\begin{proposition}
\label{proposition3}
(i)  The Hamiltonian functions given by
\bea
h_{z,1}=  \frac 1 {\xx-\yy} \, ,  \qquad
 h_{z,2}= \frac 12 \shc \bigl(\tfrac{2z} {\xx-\yy}\bigr) \biggl(   \frac{ \xx+\yy}{ \xx-\yy} \biggr) , \qquad h_{z,3}=   \frac{   \shc^2 \bigl(\frac{2z} {\xx-\yy}\bigr)( \xx+\yy)^2 -( \xx-\yy)^2 }{ 4   \shc \bigl(\frac{2z} {\xx-\yy}\bigr)  (\xx-\yy) }  \, ,
\nonumber
\eea
satisfy the  commutation relations (\ref{gb2})   with respect to the symplectic form  $\omega$ (\ref{ee})   and define  the deformed Poisson algebra  $C^\infty({\cal H}_{z,\omega}^{\rm {2R}*})$.\\
(ii) Their corresponding    deformed   vector fields   turn out to be
\begin{eqnarray}
&& {\bf X}_{z,1}   =  \frac{\partial}{\partial \xx}+  \frac{\partial}{\partial \yy},\nonumber\\[2pt]
&& {\bf X}_{z,2}=\frac 12  (\xx+\yy)  \cosh \bigl(\tfrac{2z} {\xx-\yy}\bigr)     \left(  \frac{\partial}{\partial \xx}+  \frac{\partial}{\partial \yy} \right)
+\frac 12  (\xx-\yy)  \shc \bigl(\tfrac{2z} {\xx-\yy}\bigr)     \left(  \frac{\partial}{\partial \xx}-  \frac{\partial}{\partial \yy} \right), \nonumber\\
&& {\bf X}_{z,3}   =  \frac 14 \left[  (\xx+\yy)^2+\frac{ (\xx-\yy)^2 }{ \shc^2 \bigl(\tfrac{2z} {\xx-\yy}\bigr)}  \right] \cosh \bigl(\tfrac{2z} {\xx-\yy}\bigr)     \left(  \frac{\partial}{\partial \xx}+  \frac{\partial}{\partial \yy} \right)
+\frac 12  (\xx^2-\yy^2)  \shc \bigl(\tfrac{2z} {\xx-\yy}\bigr)     \left(  \frac{\partial}{\partial \xx}-  \frac{\partial}{\partial \yy} \right) ,\nonumber
\end{eqnarray}
which fulfill 
\bea
&& [{\bf X}_{z,1},{\bf X}_{z,2}]=\cosh \bigl(\tfrac{2z} {\xx-\yy}\bigr)  {\bf X}_{z,1},\qquad [{\bf X}_{z,1},{\bf X}_{z,3}]=2 {\bf X}_{z,2},\nonumber\\[2pt] 
&& [{\bf X}_{z,2},{\bf X}_{z,3}]=\cosh \bigl(\tfrac{2z} {\xx-\yy}\bigr)   {\bf X}_{z,3}- z^2\left[ 1  - \biggl(\frac{\xx+\yy}{\xx-\yy} \biggr)^2\!  \shc^2 \bigl(\tfrac{2z} {\xx-\yy}\bigr)  \right]  {\bf X}_{z,1}. 
\nonumber
\eea
\end{proposition}

 The deformed counterpart of the coupled Ricatti Lie system (\ref{eb}) and of the   LH one (\ref{ee2}) is defined by
 \be
{\bf X}_z:=a_0(t){\bf X}_{z,1}+a_1(t){\bf X}_{z,2}+a_2(t){\bf X}_{z,3},\qquad h_z:=a_0(t)h_{z,1}+a_1(t)h_{z,2}+a_2(t)h_{z,3} .
\label{el}
\ee
And the $t$-independent constants of motion are $F_z=-1/4$ and
\bea
&&  F_z^{(2)}=\frac{   \eee^{-\frac{2z}{\xx_1-\yy_1}} \eee^{\frac{2z}{\xx_2-\yy_2}}   }{4(\xx_1-\yy_1)(\xx_2-\yy_2)} \left[   \shc \bigl(\tfrac{2z} {\xx_1-\yy_1}\bigr)      \shc \bigl(\tfrac{2z} {\xx_2-\yy_2}\bigr)  (\xx_1-\xx_2+\yy_1-\yy_2)^2  \right. \nonumber\\[2pt]
&&  \qquad  \left. -  \left(  \frac{  \eee^{\frac{2z}{\xx_1-\yy_1}}  (\xx_1-\yy_1) }{  \shc \bigl(\tfrac{2z} {\xx_1-\yy_1}\bigr) }   + \frac{  \eee^{-\frac{2z}{\xx_2-\yy_2}}  (\xx_2-\yy_2) }{  \shc \bigl(\tfrac{2z} {\xx_2-\yy_2}\bigr) }   \right)  \shc \bigl(\tfrac{2z} {\xx_1-\yy_1}+\tfrac{2z} {\xx_2-\yy_2}\bigr) (\xx_1+\xx_2-\yy_1-\yy_2)      \right] .
 \label{em}\nonumber
\eea
Therefore, the  deformation of the system (\ref{ea}) is determined  by ${\bf X}_z$ (\ref{el}).  Note that the resulting system  presents a strong interaction amongst the variables $(u,v)$ through $z$,
which goes far beyond the initial (naive) coupling corresponding to set the same $t$-dependent parameters $a_i(t)$ in both one-dimensional Riccati equations; namely 
 \bea
&& \frac{{\rm d} \xx}{{\rm d} t}=a_0(t)+\frac{a_1(t)}{2}\left[  (\xx+\yy)  \cosh \bigl(\tfrac{2z} {\xx-\yy}\bigr)    + (\xx-\yy)  \shc \bigl(\tfrac{2z} {\xx-\yy}\bigr)     \right]
  \nonumber\\[2pt]
&&\qquad\quad +\frac{a_2(t)}4 \left[    \left(  (\xx+\yy)^2+\frac{ (\xx-\yy)^2 }{ \shc^2 \bigl(\tfrac{2z} {\xx-\yy}\bigr)}  \right) \cosh \bigl(\tfrac{2z} {\xx-\yy}\bigr)       + 2 (\xx^2-\yy^2)  \shc \bigl(\tfrac{2z} {\xx-\yy}\bigr)    \right]   , \nonumber\\[2pt]
&& \frac{{\rm d} \yy}{{\rm d} t}=a_0(t)+\frac{a_1(t)}{2}\left[  (\xx+\yy)  \cosh \bigl(\tfrac{2z} {\xx-\yy}\bigr)   - (\xx-\yy)  \shc \bigl(\tfrac{2z} {\xx-\yy}\bigr)     \right]
  \nonumber\\[2pt]
&&\qquad\quad +\frac{a_2(t)}4 \left[    \left(  (\xx+\yy)^2+\frac{ (\xx-\yy)^2 }{ \shc^2 \bigl(\tfrac{2z} {\xx-\yy}\bigr)}  \right) \cosh \bigl(\tfrac{2z} {\xx-\yy}\bigr)       - 2 (\xx^2-\yy^2)  \shc \bigl(\tfrac{2z} {\xx-\yy}\bigr)    \right]   .
\label{en}\nonumber
\eea



\section{Conclusions}

 In this work, the notion of Poisson--Hopf deformation of  LH systems
has been proposed. This  framework differs radically from
other approaches to the LH systems theory~\cite{PW,Dissertations, CLS13, CGL10, BCHLS13Ham}, as our resulting deformations do not formally correspond to LH systems, but to an extended notion of   them that requires a (non-trivial)  Hopf structure and is related with the non-deformed LH system by means of a limiting process in which the deformation parameter $z$ vanishes.  Moreover, the introduction of Poisson--Hopf structures  allows for the generalization of the type of systems under inspection, since the finite-dimensional Vessiot--Guldberg Lie algebra is replaced by an involutive distribution in the Stefan--Sussman sense. 

This framework has been illustrated via the Poisson analogue of the non-standard quantum deformation of 
$\mathfrak{sl}(2)$, and deformations of physically relevant LH systems such as the oscillator system, as well as the complex and coupled Riccati equations have been presented. In the former case the deformation can be interpreted as the transformation of the initial system into a new one
possessing a position-dependent mass, hence suggesting an alternative approach to the latter type of systems that presents an ample potential of applications. In particular, the Schr\"odinger problem for position-dependent mass Hamiltonians is directly connected with the quantum dynamics of charge carriers in semiconductor heterostructures and nanostructures (see, for instance,~\cite{Roos, Bastard, qDWW}). In this respect, it is worth remarking that the standard or Drinfel'd--Jimbo deformation of 
$\mathfrak{sl}(2)$ would not lead to an oscillator with a position-dependent mass since, in that case, the deformation function would be $\!\shc(z q p)$ instead of $\!\shc(z q^2)$; this     can clearly be  seen in the corresponding symplectic realization given in~\cite{AR}. This fact explains that, in order to illustrate our approach, we have chosen the non-standard deformation of $\mathfrak{sl}(2)$  due to its physical applications.
 In spite of this, the Drinfel'd--Jimbo deformation would provide another deformation for the MP and Riccati  equations which would be 
non-equivalent  to the ones here studied.

There are still many questions to be analyzed in detail. Since the formalism here presented is applicable in a more wide context, with other types of Hopf algebra deformations and dealing with higher-dimensional Vessiot--Guldberg Lie algebras, this would lead to a richer spectrum of properties for the deformed systems that deserve further investigation. For instance, the deformed LH systems studied in this work are such that the distribution spanned by the deformed 
vector fields is the same as the initial one. As it has been observed previously, this constraint could not be preserved for generic Poisson--Hopf algebra deformations of LH systems defined 
on more general manifolds.

An important question to be addressed is whether this approach can provide an effective procedure to  derive a  deformed analogue of superposition principles for deformed LH systems. Also, it would be interesting to know
whether such a description is simultaneously applicable to the various non-equivalent deformations, like an extrapolation of the notion of Lie algebra contraction to Lie systems. Another open problem worthy to be considered  is the possibility of getting a unified description of such systems in terms of a certain amount of fixed  `elementary' systems, thus implying a first rough systematization of LH-related systems from a more general perspective than that of finite-dimensional Lie algebras. Work in these directions is currently in progress.


\section*{Appendix. The hyperbolic sinc function}
\setcounter{equation}{0}
\renewcommand{\theequation}{A.\arabic{equation}}

\begin{figure}[t]
\begin{center}
\includegraphics[height=6.0cm]{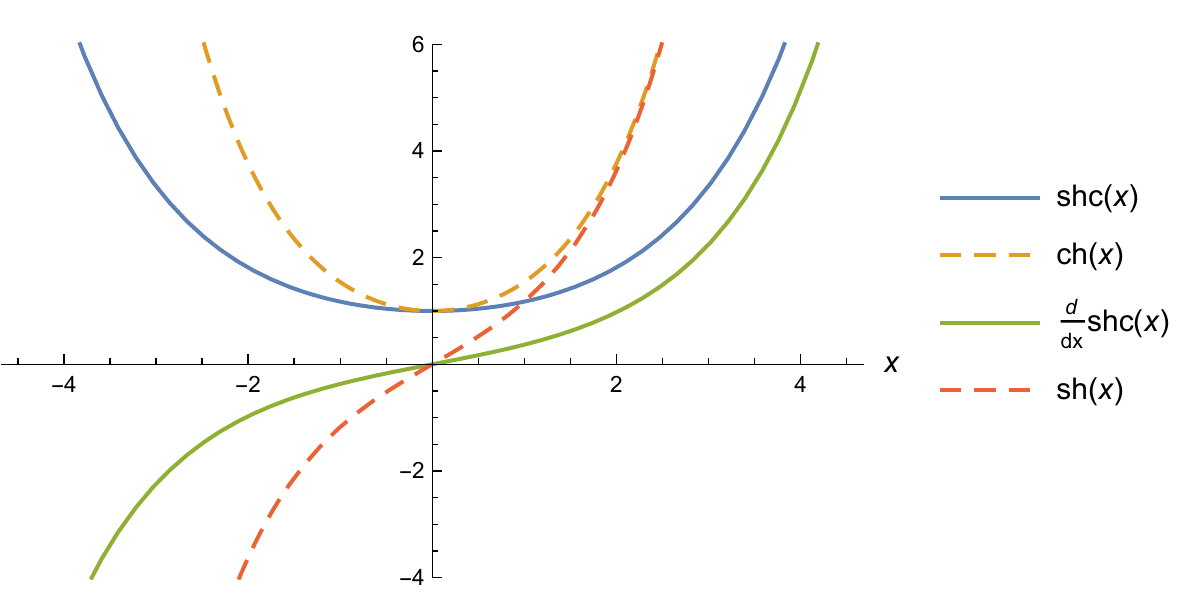}
\caption{\small The hyperbolic sinc function versus the hyperbolic cosine function and the derivative of the former versus the hyperbolic sine function.}
 \label{fig3}
\end{center}
\end{figure}

   The hyperbolic counterpart of the  well-known sinc  function is defined by
\be
\shc( x):=  \left\{ 
\begin{array}{ll}
\frac{\sinh (x)}{x}, &\mbox{for}\ x\ne 0, \\
1,&\mbox{for}\  x=0.
\end{array}
 \right. 
 \nonumber
\ee
 The power series around $x=0$ reads
 \be
 \shc( x) = \sum_{n=0}^\infty \frac{x^{2n}}{(2n+1)!} \, .
 \nonumber
 \ee
And its derivative is given by
\be
\frac{\rm d}{{\rm d}x} \shc(x)= \frac{\cosh (x)}{x}- \frac{\sinh (x)}{x^2} =  \frac {\cosh( x) - \shc(x)}{x} \, .
\nonumber
\ee
 Hence the behaviour of $\shc( x)$ and its derivative remind that of the hyperbolic cosine and sine functions, respectively.  
  We represent them  in figure~\ref{fig3}.


  A novel relationship of the  $\shc$ function  (and also of the $\sinc$ one) with Lie systems can be established   by considering the following second-order ordinary differential equation
  \be
t\,  \frac{\dd^2 x}{\dd t^2}+ 2 \,  \frac{\dd x}{\dd t}-\eta^2  t\, x =0 ,
\label{diff}
  \ee
  where $\eta$ is a non-zero real parameter. Its general solution can be written as
  \be
  x(t)=A \shc(\eta t)+B\, \frac{\cosh(\eta t)}{t} \,  ,\qquad A,B\in \mathbb{R}.
  \nonumber
  \ee
Notice that if we set $\eta=i \lambda$ with $\lambda\in \mathbb{R}^\ast$ we recover the known result for the sinc function:
  \be
  t\,  \frac{\dd^2 x}{\dd t^2}+ 2 \,  \frac{\dd x}{\dd t}+\lambda^2  t\, x =0 , \qquad x(t)=A \sinc(\lambda t)+B\, \frac{\cos(\lambda t)}{t} \,  .
  \label{Ad}
  \ee
   Next the differential equation (\ref{diff})  can be written as a system of two first-order differential equations by setting $y={\dd x}/{\dd t}$, namely
\be
\frac{\dd x}{\dd t}=y,\qquad \frac{\dd y}{\dd t}=- \frac{2}{t}\, y+ \eta^2 x.
\label{eqLie}
\nonumber
\ee
Remarkably enough,  these equations  determine a Lie system  with  associated  $t$-dependent vector field 
\be
{\bf X}=- \frac 2 t\, {\bf X}_1+ {\bf X}_2 + \eta^2 {\bf X}_3,
\label{Aa}
\ee
where
\begin{equation} 
{\bf X}_1=y\frac{\partial}{\partial y},\qquad {\bf X}_2= y\frac{\partial}{\partial x} ,\qquad {\bf X}_3=  x\frac{\partial}{\partial y} ,\qquad {\bf X}_4=x \frac{\partial}{\partial x}+ y\frac{\partial}{\partial y},
\nonumber
\end{equation}
   fulfill the commutation relations
\begin{equation} 
[{\bf X}_1,{\bf X}_2]={\bf X}_2,\qquad [{\bf X}_1,{\bf X}_3]=-{\bf X}_3,\qquad [{\bf X}_2,{\bf X}_3]=2 {\bf X}_1- {\bf X}_4 ,\qquad [{\bf X}_4, \, \cdot \, ]=0 .
\nonumber
\end{equation}
Hence, these  vector fields span a Vessiot--Guldberg Lie algebra  $V$   isomorphic to $\mathfrak{gl}(2)$ with domain   $\mathbb R^2_{x\ne 0}$. 
In fact,  $V$ is diffeomorphic to the class ${\rm I}_7 \simeq \mathfrak{gl}(2)$ of the classification given in~\cite{BBHLS}. The diffemorphism can be explictly performed by means of the change of variables
$u=y/x$ and $v=1/x$,  leading to the vector fields of class ${\rm I}_7$ with domain $\mathbb R^2_{v\ne 0}$ given in~\cite{BBHLS}
$$
{\bf X}_1= u\frac{\partial}{\partial u},\qquad  {\bf X}_2= -u^2\frac{\partial}{\partial u}- u  v \frac{\partial}{\partial v},\qquad
{\bf X}_3=  \frac{\partial}{\partial u},\qquad  {\bf X}_4= - v \frac{\partial}{\partial v} .
$$
Therefore ${\bf X}$ (\ref{Aa}) is a Lie system but not a  LH one since there does not exist any compatible symplectic form  satisfying  (\ref{der}) for  class ${\rm I}_7$  as shown in~\cite{BBHLS}.

Finally, we   point out that the very same result follows  by starting from the differential equation (\ref{Ad}) associated with the sinc function.


\section*{Acknowledgments}

\small

A.B.~and F.J.H.~have been partially supported by Ministerio de Econom\'{i}a y Competitividad (MINECO, Spain) under grants MTM2013-43820-P and   MTM2016-79639-P (AEI/FEDER, UE), and by Junta de Castilla y Le\'on (Spain) under grants BU278U14 and VA057U16.  The research of R.C.S.~was partially   supported by grant MTM2016-79422-P  (AEI/FEDER, EU).
E.F.S.~acknowledges a fellowship (grant CT45/15-CT46/15) supported by the  Universidad Complutense de Madrid.   J.~de L.~acknowledges funding from the Polish National Science Centre under grant HARMONIA 2016/22/M/ST1/00542. 


\end{document}